\documentclass[epj]{svjour}
\usepackage{url}
\usepackage{hyperref}
\usepackage{xspace}
\usepackage{graphicx}
\usepackage{amssymb}
\usepackage{multirow}
\usepackage{array}
\usepackage[usenames,dvipsnames]{color}
\usepackage[pagewise]{lineno}
\usepackage{graphics}
\usepackage{color}

\def\npb{{\em Nucl. Phys. {\bf B} }}

\begin{document}

\title{Speed of Sound in Hadronic matter using Non-extensive Tsallis Statistics}

\author{\large Arvind Khuntia$^{1}$, \large Pragati Sahoo$^{1}$, \large Prakhar Garg$^{1,}$\thanks{\emph {Present address}: Department of Physics and Astronomy, Stony Brook University, SUNY, Stony Brook, New York 11794-3800, USA}, \large Raghunath Sahoo$^{1,}$\thanks {email: Raghunath.Sahoo@cern.ch (corresponding author)}, \and Jean Cleymans$^{2}$} 



\institute{Discipline of Physics, School of Basic Science, Indian Institute of Technology Indore, Khandwa Road, Simrol, M.P. India-453552 \label{addr1}
           \and
           UCT-CERN Research Centre and Department of Physics,
University of Cape Town, Rondebosch 7701, South Africa\label{addr2}}
\date{Received: date / Revised version: \today}

\abstract{
The speed of sound ($c_s$) is studied to understand the hydrodynamical evolution of the matter created in
heavy-ion collisions. The quark-gluon plasma (QGP) formed in heavy-ion collisions evolves from an initial QGP to the hadronic phase via a possible mixed phase. Due to the system expansion in a first order phase transition
scenario, the speed of sound reduces to zero as the specific heat diverges. We study the speed of sound for systems, which deviate from a thermalized Boltzmann distribution using non-extensive Tsallis statistics. In the present work, we calculate the speed of sound as a function
of temperature for different $q$-values for a hadron resonance gas. 
We observe a similar mass cut-off behaviour in the non-extensive case for $c^{2}_s$ by including heavier particles, as is observed in the case of a hadron resonance gas following equilibrium statistics. Also, we explicitly show that the temperature where the mass cut-off starts varies with the $q$-parameter which hints at a relation between the degree of non-equilibrium and the limiting temperature of the system. It is shown that for values of $q$ above approximately 1.13 all
criticality disappears in the speed of sound, i.e. the decrease in the
value of the speed of sound, observed at lower values of $q$, disappears completely.
\PACS{
{25.75.Dw,} 
{25.75.Nq,}
{24.10.Pa,}
{51.30.+i}
}
}
\authorrunning{Arvind Khuntia et al.} 

\maketitle
\section{Introduction}
In high energy hadronic and nuclear collisions, the space-time evolution of the created hot and dense matter is controlled by the initial energy density and the created temperature. Because of the very high initial pressure, the system expansion takes place through the decrease of its temperature and energy density. The change in pressure with energy density is related to a physical observable called the speed of sound inside the system, which can be used to probe the degrees of freedom, as it is related to the equation of state (EoS) of the system. This also helps in looking for any critical behaviour the system could undergo during its space-time evolution. The study of the speed of sound in hadronic matter hence becomes a subject of paramount importance in heavy-ion collisions. There have been several experimental and theoretical studies in this direction \cite{Chojnacki:2007jc,Steinheimer:2012bp,Tawfik:2012ty,Mohanty:2003va,Epele:1986qz,Plumer:1984aw,Gavai:2004se,castorina}. The speed of sound has also been discussed in a slightly different context in Ref.~\cite{Deppman:2012qt}.
It should be noted here that the speed of sound  in a medium can be derived from Euler's law
as applied to a continuous medium without having to use thermodynamics \cite{Landau-L}.
The temperature dependence of the speed of sound in a medium is well established but the effect of temperature fluctuations is less explored, particularly in the case of high-energy collisions, where the non-extensive parameter $q$ lies in $1 < q < 1.2$ \cite{Beck:2000nz}. As $q=1$ corresponds to an equilibrated system described by Boltzmann-Gibbs statistics, the systems with small values of $(q-1)$ are assumed to undergo an infinitesimal perturbation and tend to approach equilibrium. In non-extensive statistics, the Tsallis parameter ($q$) is related to the temperature fluctuations~\cite{Wilk:1999dr} and therefore we estimate the speed of sound using Tsallis statistics in hadronic medium, and study it as a function of $q$, for systems undergoing small perturbations. 
Further, it has been observed that the transverse momentum spectra of the secondaries created in high energy $p+p(\bar{p})$ \cite{Deppman:2012qt,Deppman:2012us,Sena:2012ds,Azmi:2014dwa,Wong:2013sca,Cleymans:2013rfq,Wong:2012zr,Cleymans:2011in}, $e^++e^-$ collisions \cite{Bediaga:1999hv,Urmossy:2011xk} are better described by the non-extensive Tsallis statistics \cite{Tsallis:1987eu}. In addition, non-extensive statistics with radial flow successfully describes the spectra at intermediate $p_{\rm T}$ in heavy-ion collisions \cite{Tang:2008ud,Bhattacharyya:2015hya,De:2007zza,De:2014dna,De:2014pqa}. Tsallis non-extensive statistics in Boltzmann Transport Equation with relaxation time approximation successfully describes the nuclear modification factor of heavy flavors in heavy-ion collisions at RHIC and LHC energies \cite{Tripathy:2016hlg}. 
Tsallis distributions have also been used to study the conserved number susceptibilities in heavy-ion collisions \cite{Mishra:2013qoa}. The present paper elucidates on the speed of sound in the framework of Tsallis non-extensive statistics taking a hadron resonance gas.
 
 
 \par
 In this paper, first we discuss the thermodynamics of a hadron
 resonance gas in Section 2, taking a grand canonical partition function
 and using Boltzmann-Gibbs statistics. In Section 3, we derive the
 necessary formalism for calculating the speed of sound using Tsallis non-extensive
 statistics. In addition, we discuss the speed of sound in a hadron
 resonance gas in the framework of non-extensive statistics with
 different values of $q$ and mass cut-off in Section 3. We summarize
 the results and conclude in Section 4.

\section {Thermodynamics of a Hadron Resonance Gas}
As we make a direct comparison of the thermodynamic quantities with those of an ideal pion gas, obeying Boltzmann statistics, for completeness we derive the necessary tools following Ref. \cite{castorina}. The grand canonical partition function for an ideal pion gas having three degrees of freedom is given by
\begin{eqnarray}
Z(T,V)&=&\sum _{N} \frac{1}{N!} \left[\frac{3V}{(2\pi)^{3}} \int_{0}^{\infty} d^3p \right.\nonumber\\
&&\left.\exp{\frac{-\sqrt{p^2+{m_{0}}^2}}{T} }\right]^N,
\end{eqnarray}
where $V$ is the system volume at temperature $T$ and $m_0$ is the mass of a pion. The logarithm of the partition function gives
\begin{eqnarray}
\ln Z(T,V)=\frac{3VT{m_{0}^2}}{2\pi^2} ~ K_{2} \left(m_{0}/T\right),
\end{eqnarray}

One obtains the basic thermodynamic quantities like, pressure ($P_0$), energy density ($\epsilon_0$) and entropy density ($s_0$) using the above partition function as

\begin{eqnarray}
P_{0}(T)&=& T \left(\frac {\partial \ln Z}{\partial V}\right)_{T}
= \frac {3{m_{0}}^2 T^2}{2\pi^2} K_2\left( m_{0}/T\right),
\label{p0}
\end{eqnarray}

\begin{eqnarray}
\label{epsilon_pion}
&&\epsilon _{0}(T)=T^2~ \left(\frac{ \partial \ln Z}{\partial T}\right)\nonumber\\
&=&\frac {3{m_{0}}^2 T^2}{2\pi^2}\left[3K_2\left(m_0/T\right)+\left(m_0/T\right) K_1\left(m_0/T\right)\right] , 
\label{e0}
\end{eqnarray}
and
\begin{eqnarray}
&& s_0(T)=\frac{\epsilon _{0}(T)+P_0(T)}{T}\nonumber\\
&=&\frac {3{m_{0}}^2 T}{2\pi^2}\left[4K_2\left(m_0/T\right)+\left(m_0/T\right) K_1\left(m_0/T\right)\right],
\label{s0}
\end{eqnarray}
where $K_1$ and $K_2$ are the well-known modified Bessel functions of second kind. 

For an ideal gas with zero chemical potential, the temperature dependent speed of sound, $c_s(T)$, is given by
\begin{equation} 
c_s^2(T) = \left(\frac{\partial P}{\partial\epsilon}\right)_V ~=~ \frac{s(T)}{C_V(T)},
\label{velOfSound}
\end{equation}
where
\begin{equation}
s= \left(\frac{\partial P}{\partial T}\right)_V
\label{entropy}
\end{equation}
is the entropy density and 
\begin{equation}
C_V(T) = \left(\frac{\partial \epsilon}{\partial T}\right)_V
\label{cv}
\end{equation} 
is the specific heat at constant volume. The specific heat at constant volume for an ideal pion gas using Eqs. \ref{e0} and  \ref{cv} is given by

\begin{eqnarray}
C_V^0(T)=3s_0(T)+\frac{3m_0^4}{2\pi^2T} K_2(m_0/T)
\end{eqnarray}
Using Eqn. \ref{velOfSound} we get the speed of sound, $c_s$ as

\begin{eqnarray}
\frac{1}{c_s^2}-3 &=& \frac{ 3 m_0^4 K_2(m_0/T)}{2\pi^2Ts_0}\\
&=&\frac{m_0^2 K_2(m_0/T)}{4T^2K_2(m_0/T)+m_0 T K_1 (m_0/T)} ,
\end{eqnarray}
with $m_0$ is the pion mass, the above expression gives the speed of sound for an ideal pion gas.

 \section {Speed of Sound in a Physical Hadron Resonance Gas}
A hadron resonance gas consists of mesons and baryons obeying Bose-Einstein (BE) and Fermi-Dirac (FD) statistics, respectively. The Tsallis form of  the Fermi-Dirac and Bose-Einstein distributions as proposed 
in Refs.~\cite{Conroy:2010wt,turkey1,Pennini1995309,Teweldeberhan:2005wq,Conroy:2008uy,Chen200265} is
\begin{equation}
f_T(E) \equiv 
\frac{1}{\exp_q\left(\frac{E-\mu}{T}\right) \pm1}  .
\label{tsallis-fd}
\end{equation}
where the function $\exp_q(x)$ is defined as
\begin{equation}
\exp_q(x) \equiv \left\{
\begin{array}{l l}
\left[1+(q-1)x\right]^{1/(q-1)}&~~\mathrm{if}~~~x > 0 \\
\left[1+(1-q)x\right]^{1/(1-q)}&~~\mathrm{if}~~~x \leq 0 \\
\end{array} \right.
\label{tsallis-fd1}
\end{equation}
and, in the limit where $q \rightarrow 1$ it reduces to the standard exponential;
$\lim_{q\rightarrow 1}\exp_q(x)\rightarrow \exp(x)$. In the present context we have taken $\mu =0$, therefore $x\equiv E/T $ is always positive. In Eqn. \ref{tsallis-fd}, the negative sign in the denominator stands for BE and the positive stands for FD distribution.

\begin{figure}[ht!]
\includegraphics[width=0.45\textwidth]{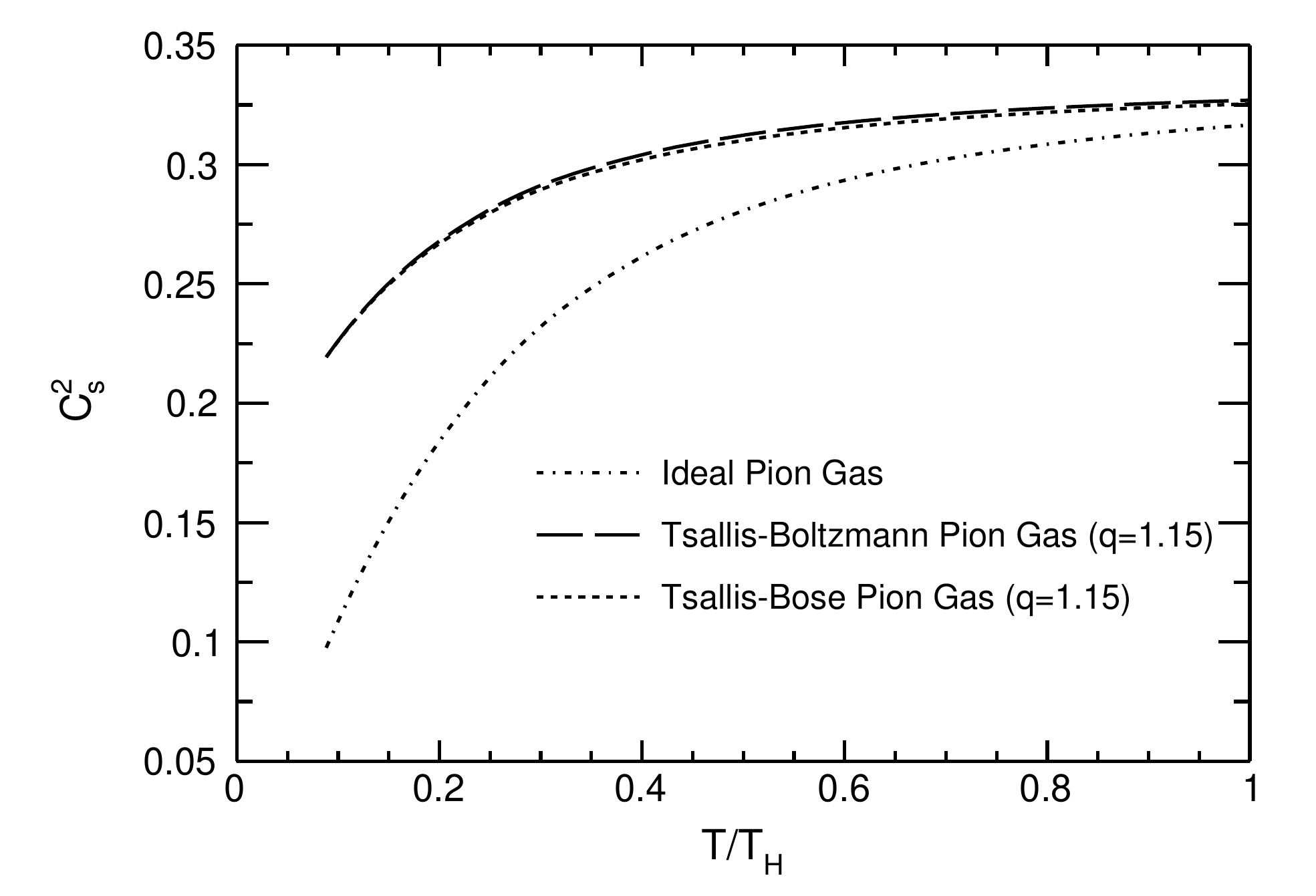}
\caption{Speed of sound as a function of scaled temperature for an ideal pion
  gas (dash-dotted line: Boltzmann distribution) and for a pion gas
  with non-extensive statistics using Boltzmann (long-dashed line) and 
Bose-Einstein statistics (dotted line), for $q=$ 1.15.}
\label{quantum}      
\end{figure}

To see the effect of quantum statistics on the speed of sound for a
pion gas, we have explicitly compared the Tsallis-BE and
Tsallis-Boltzmann distributions, which are shown in
Fig. \ref{quantum} as a function of temperature scaled with the Hagedorn limiting temperature, $T_{\rm H}$. 
It is observed that for $q=1.15$, there is no
distinction between both the distributions, {\it i.e.} between
the Tsallis-Boltzmann and the Tsallis-BE gas, and hence one considers
Tsallis-Boltzmann distributions as a good approximation for describing
the thermodynamics of a pion gas, as is done in the previous
section. However, as expected, there is a clear distinction between
equilibrium and Tsallis non-extensive statistics, for which the degree
of deviation is higher towards lower temperatures.  

\par
For a hadron resonance gas, the energy density, pressure and specific heat at constant volume in non-extensive statistics are given by

\begin{eqnarray}
~~\epsilon &=&\frac{g}{2\pi^2} \int_{0}^{\infty} dp~p^2 \sqrt{p^2+m_0^2}\nonumber\\
&&\times \left[\left[1+\frac{(q-1)\sqrt{p^2+m_0^2}}{T}\right]^{\frac{1}{q-1}} \mp1\right]^{-q}
\end{eqnarray}

\begin{eqnarray}
 ~~P&=&\frac{g}{2\pi^2} \int_{0}^{\infty} dp~p^4\frac{1}{3\sqrt{p^2+m_0^2}}\nonumber\\
&& \times\left[\left[1+\frac{(q-1)\sqrt{p^2+m_0^2}}{T}\right]^{\frac{1}{q-1}} \mp1\right]^{-q}
\end{eqnarray}
and,
\begin{eqnarray}
 ~~C_V &=& \frac{qg}{2\pi^2 T^2} \int_{0}^{\infty} dp~p^2 (p^2+m_0^2)\nonumber\\
&&\times\frac{\left[1+\frac{(q-1)\sqrt{p^2+m_0^2}}{T}\right]^{\frac{2-q}{q-1}}}
{\left[\left[1+\frac{(q-1)\sqrt{p^2+m_0^2}}{T}\right]^{\frac{1}{q-1}} \mp1\right]^{q+1}}
\nonumber\\
\end{eqnarray}
Hence the speed of sound can be calculated from these relations using Eqn. \ref{velOfSound}. 

\begin{figure}[ht!]
\includegraphics[width=0.45\textwidth]{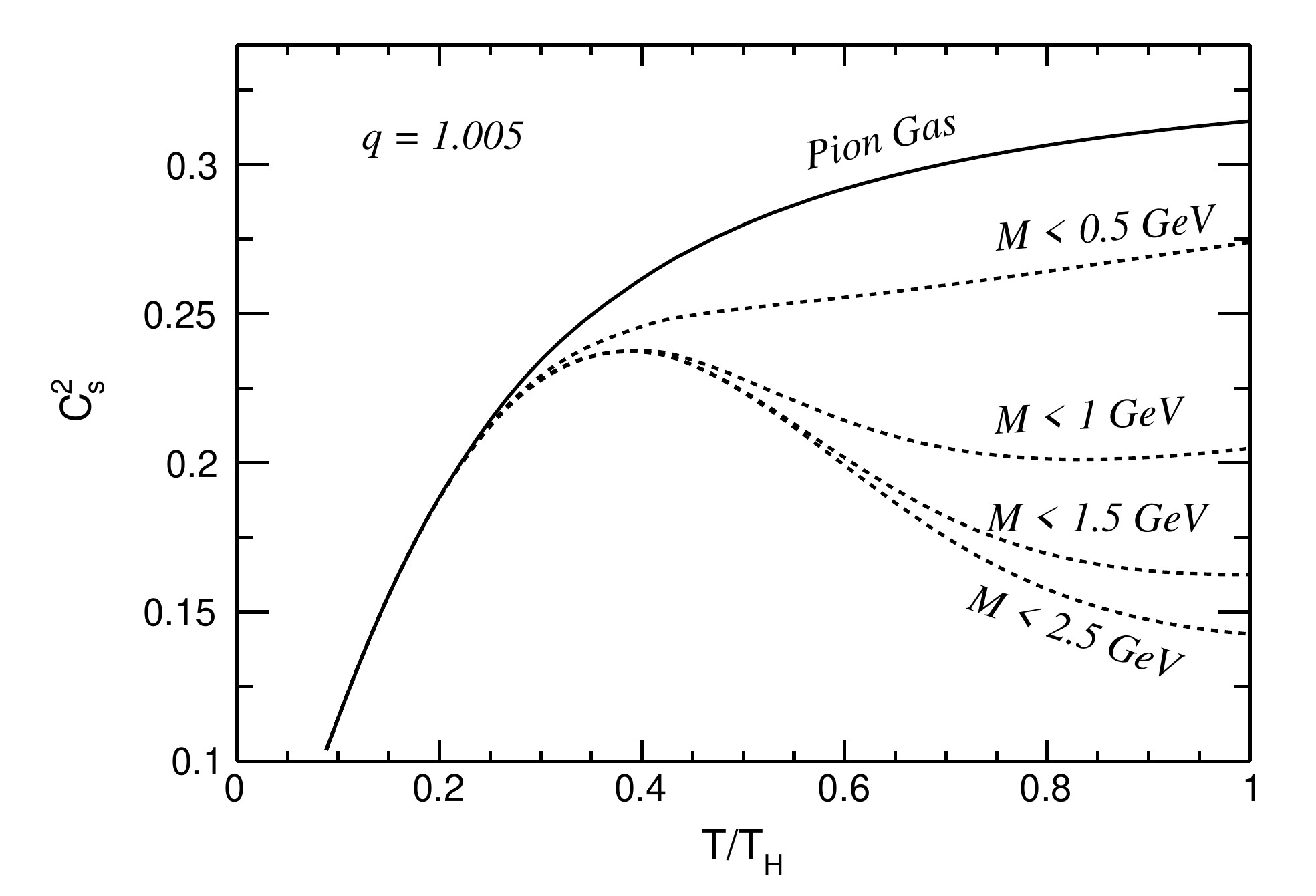}
\caption{Speed of sound for $q$=1.005 for a hadron resonance gas with different cut-off on mass.}
\label{q_1006}
\end{figure}

Figure~\ref{q_1006} shows the speed of sound for a hadron resonance gas and the effect of taking different mass cut-offs  of hadrons for a very small value of $q$ ($=$1.005). The mass cut-off $M$ is introduced as the highest mass of the resonances contributing to the hadron resonance gas. Ref. \cite{Deppman:2012qt} has also considered the case of contributions going beyond such a cut-off. It is evident from Fig.~\ref{q_1006} that in the limit of $q$$\sim$1, a hadron resonance gas in Tsallis framework  assuming quantum statistics gives identical results as shown in Ref.~\cite{castorina} for a similar system. 

An increase in the value of $q$ above one means a deviation from equilibrium statistics. 
To study the effect of higher values of $q$ on the speed of sound, we have taken $q=1.05$ and 1.15 with 
different mass cut-offs by inclusion of heavy resonances. 

Figures ~2 and 3 show that, for small values of $q$ and large mass
cut-off, the speed of sound at first increases
as a function of temperature, reaches a maximum value and then
decreases, which is indicative of a phase transition. In a first order phase transition it is expected
that the speed of sound drops
to zero at the phase transition point.
This  behavior is no longer present for larger values of $q$ as can be
seen from Figure 4 where
the speed of sound increases monotonically with temperature. This basically
is the behavior for any value of the mass cut-off.
A comparison of Figures 2 and 3 with Figure 4 opens the possibility
that there exists a
special value of $q$ where a phase transition is no longer present. 
To explore this, we have studied the square of the speed of sound as a function of 
the scaled temperature for different values of $q$, taking a mass cut-off of 2.5 GeV.
This is shown in Fig.~\ref{q_crit}.  It could be inferred from the figure that with the progressive increase of the $q$-values, 
when the system goes away from equilibrium, the speed of sound slowly decreases to a minimum value  up to
$q=1.13$. This behaviour, however, vanishes for higher values of $q$. This indicates the softening of the equation of state and a possible phase transition.  Interestingly, this critical value of $q$ is close to the value one obtains in the analysis of $p_{\rm T}$ spectra in $p+p$ collisions at high energies \cite{Azmi:2015xqa}.

In order to further explore the nature of the phase transition, we have studied the energy density ($\epsilon$) scaled with $T^4$, which is a measure of the degrees of freedom
of the system, as a function of the temperature, taking the same mass cut-off for the physical resonance gas. This is shown in 
Fig.~\ref{E_crit}. With the increase of the $q$-values,  $\epsilon/T^4$ increases faster, showing a consistent 
sharp increase in its value for some temperatures, which may indicate a possible transition to higher degrees of 
freedom. Then after reaching a maximum value for all $q$-values, it starts decreasing for higher temperature and 
goes to a saturation region. This behavior can be understood as follows. As the temperature increases more and more heavy resonances start 
contributing to the energy density and hence one notices a strong increase in the energy density. 
At infinite temperature, the system starts behaving like a gas of massless particles and $\epsilon/T^4$ ultimately goes to a constant value.
The rapid increase with the parameter $q$ occurs because the distribution deviates more and more from a Boltzmann
distribution and the tails at large momentum contribute more and more as $q$ increases.  
For higher $q$-values 
the critical behaviour of $c_{s}^{2}$ diminishes by including heavy resonances.
\begin{figure}[ht!]
\includegraphics[width=0.45\textwidth]{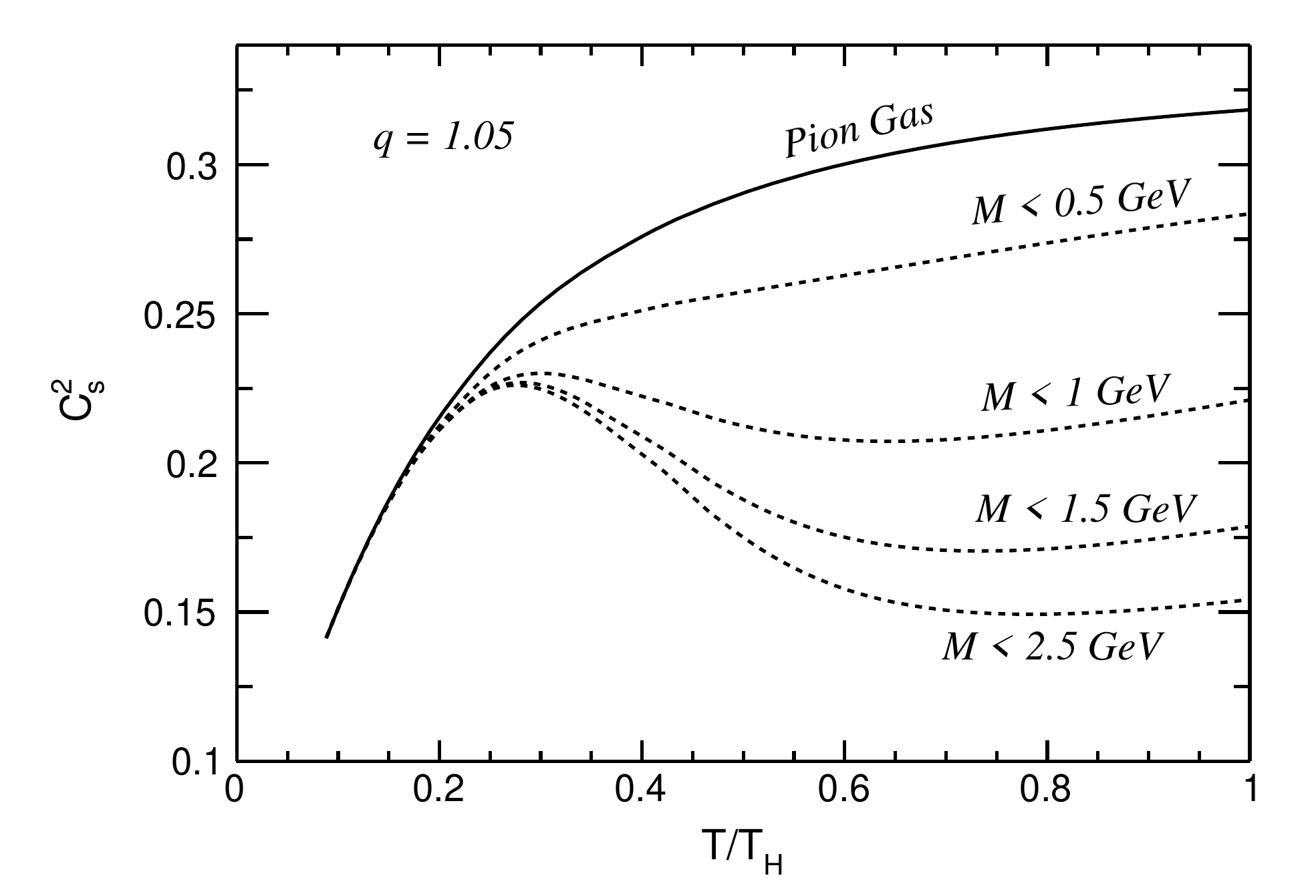}
\caption{Speed of sound for $q$=1.05 for a hadron resonance gas with different cut-off on mass.}
\label{q_105}
\end{figure}

\begin{figure}[ht!]
\includegraphics[width=0.45\textwidth]{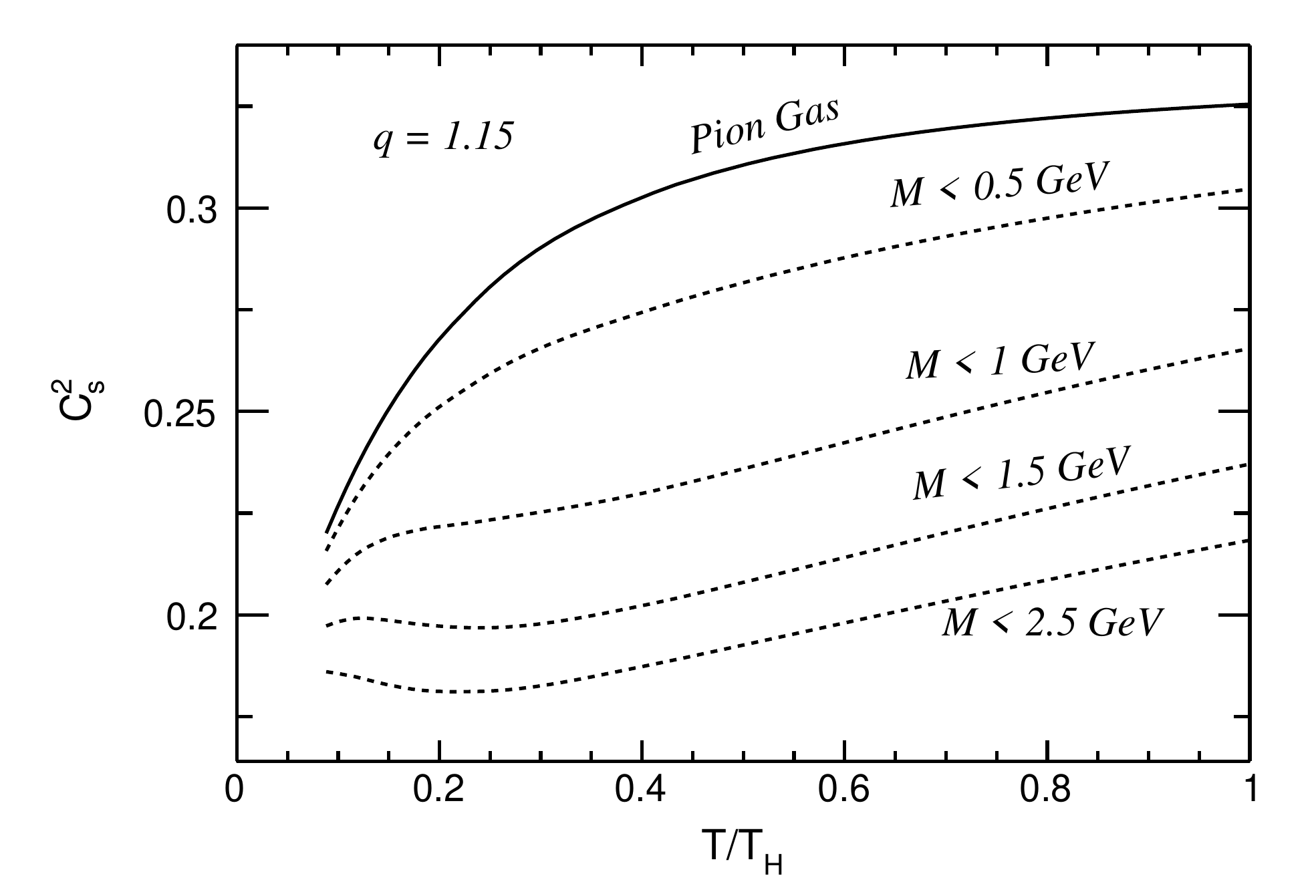}
\caption{Speed of sound for $q$=1.15 for a hadron resonance gas with different cut-off on mass.}
\label{q_115}
\end{figure}

\begin{figure}[ht!]
\includegraphics[width=0.45\textwidth]{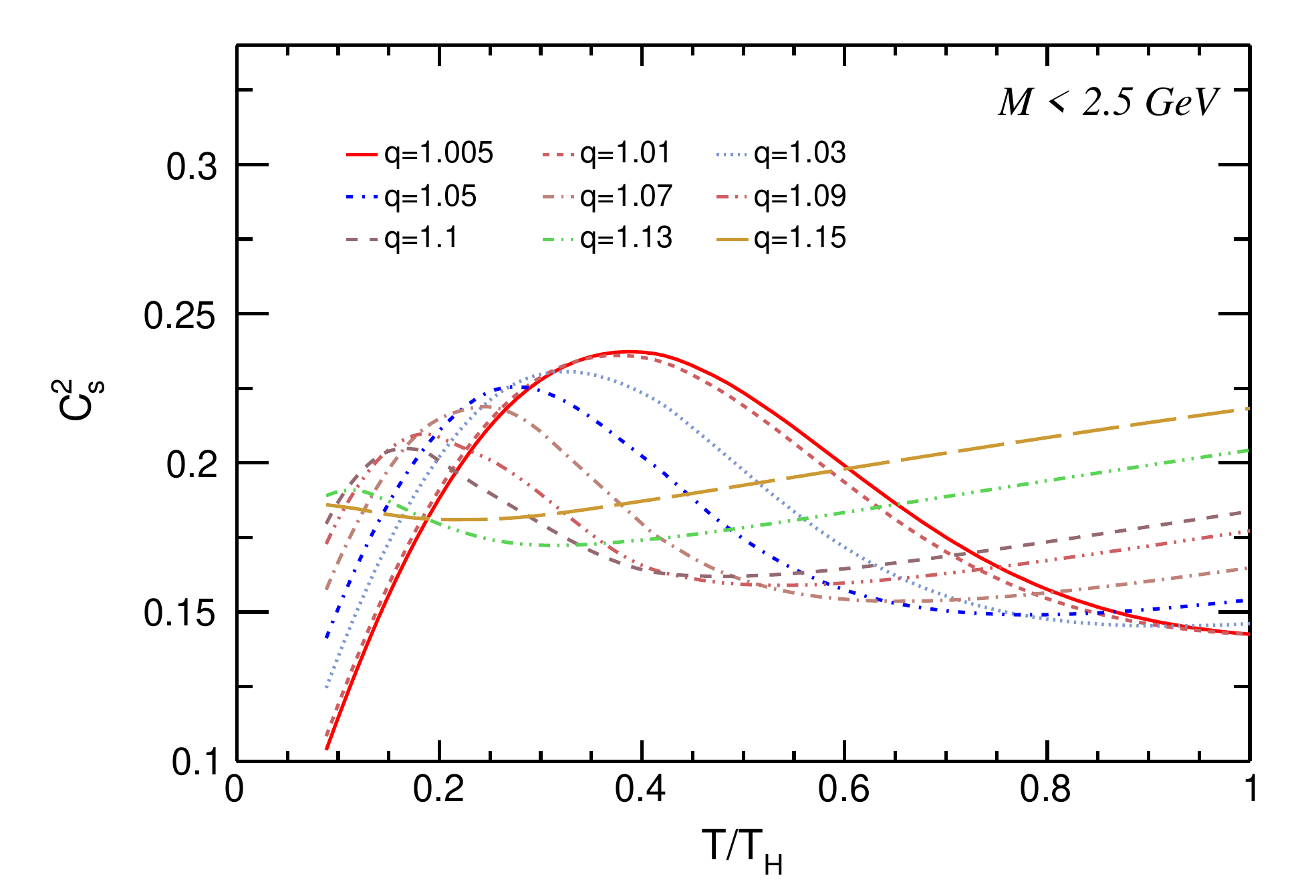}
\caption{(Color online) Speed of sound for different values of the non-extensivity parameter, $q$, for a mass cut-off of $M < 2.5$ GeV.}
\label{q_crit}
\end{figure}

\begin{figure}[ht!]
\includegraphics[width=0.45\textwidth]{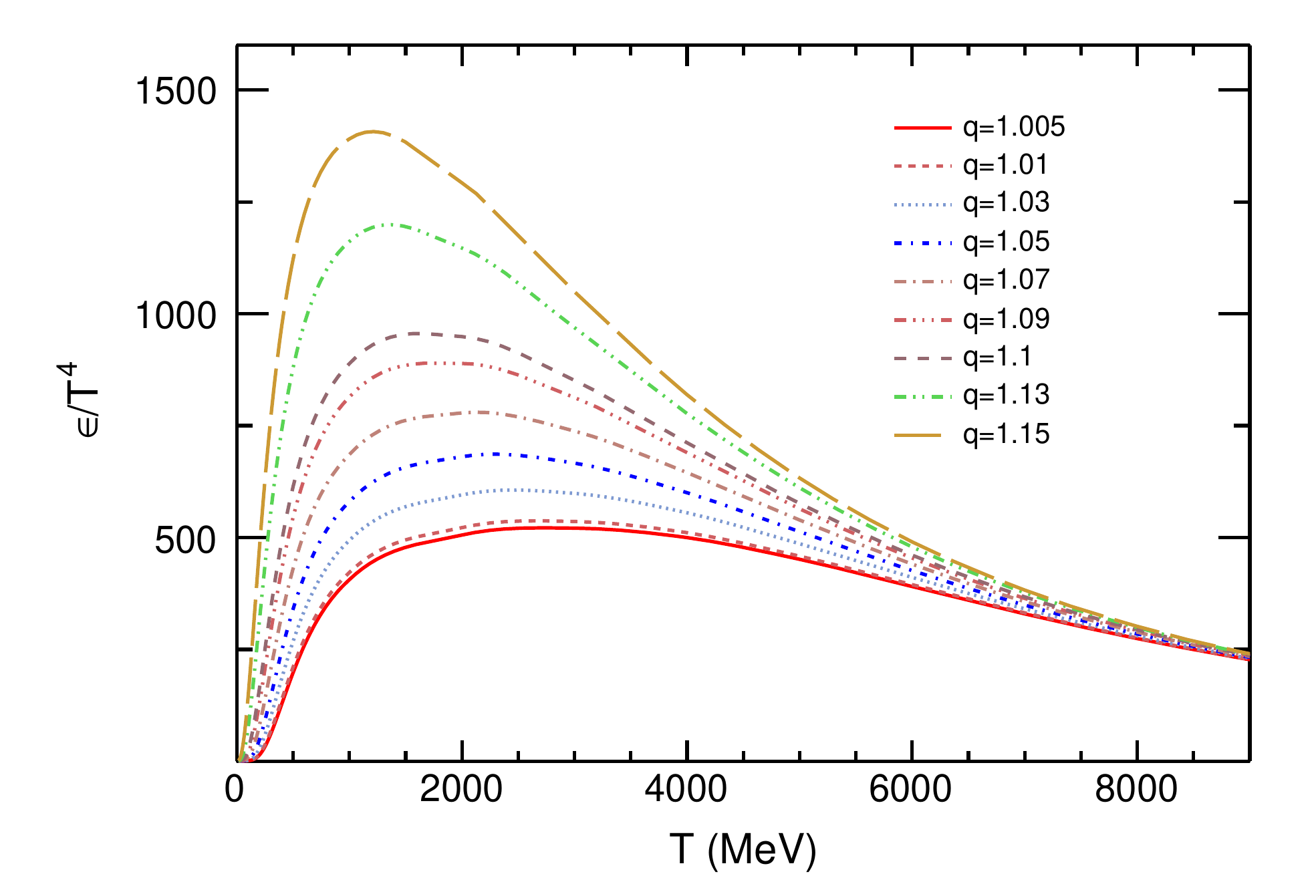}
\caption{(Color online) Energy density scaled to $T^4$ (representing the degrees of freedom of the system) for different values of the non-extensivity parameter, $q$, for a mass cut-off of $M < 2.5$ GeV.}
\label{E_crit}
\end{figure}

\begin{figure}
\includegraphics[width=0.45\textwidth]{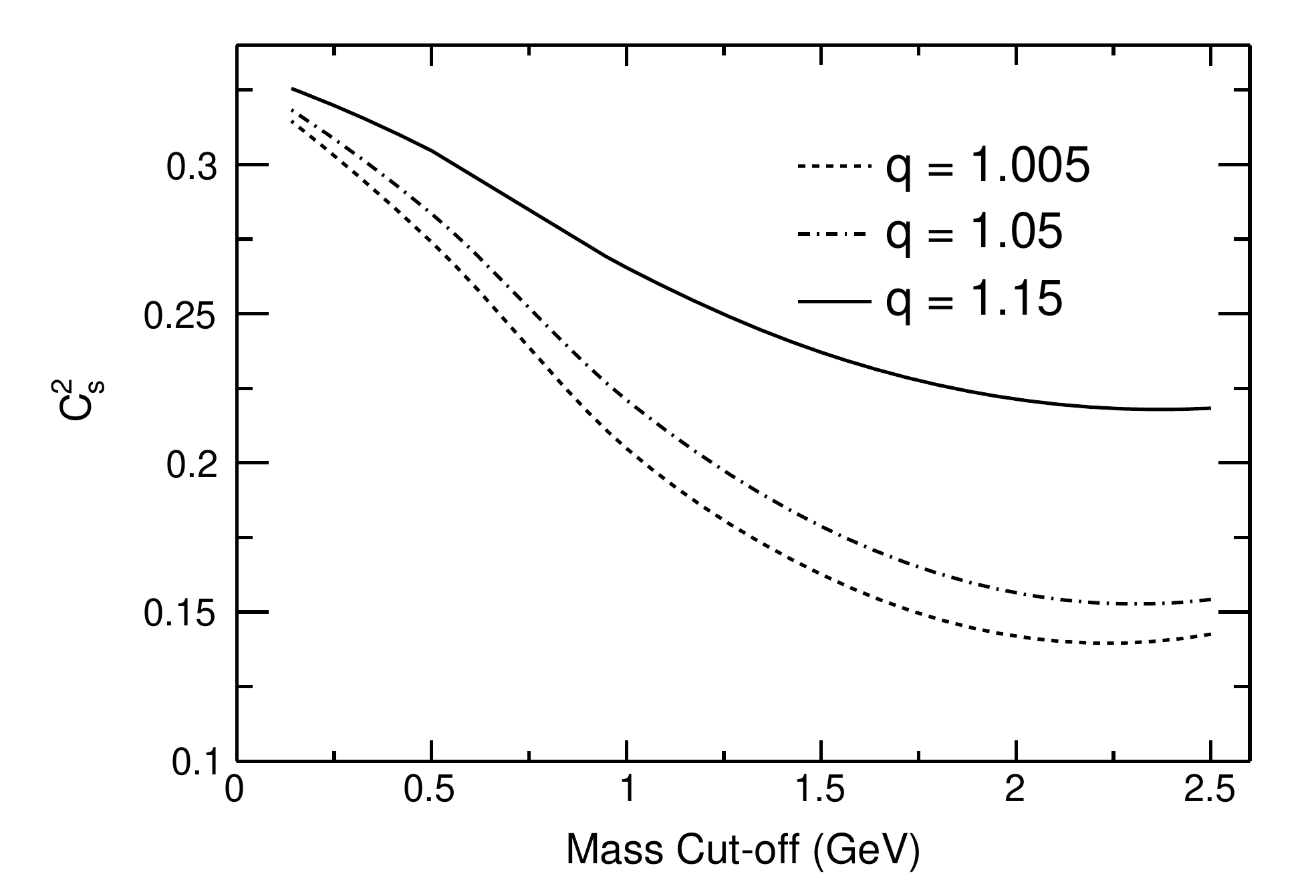}
\caption{Speed of sound as a function of upper mass cut-off for a hadron resonance gas at T =170 MeV, with different values of the parameter $q$.}
\label{fig:7}
\end{figure}

The speed of sound as a function of upper mass cut-off for a hadron resonance gas at temperature, $T$ = 170 MeV, with different $q$-values is shown in Fig.~\ref{fig:7}.  We choose a limiting temperature, $T_H \sim T_c$ = 170 MeV motivated by the lattice QCD calculations \cite{Karsch:2000kv} and also the Hagedorn limiting temperature \cite{hagedorn1,hagedorn2,hagedorn3}, where $T_H$ is the Hagedorn limiting temperature and $T_c$ is the critical temperature for a deconfinement transition. This shows a monotonic decrease of speed of sound with the inclusion of higher resonances to the system and a systematic trend in the increase of the speed of sound is seen with the increase in $q$-value of the system. 

\section{Summary and Conclusion}
The speed of sound in a hadronic medium using non-extensive statistics is calculated for a hadron resonance gas, as the Tsallis statistics allows for the exploration of systems which are away from thermal equilibrium. 
Although the speed of sound has been estimated previously as a function of the
temperature in a hadronic medium by several authors, it has not been studied for
a hadron resonance gas with temperature fluctuations. Tsallis parameter ``$q$" is related to the temperature fluctuation in a thermodynamic system. We study $c_{s}^{2}$ as a function of the non-extensive parameter $q$ in order to explore the effect of temperature fluctuations.

In the present work, the effect of different mass cut-offs on $c_{s}^{2}$ by adding massive resonances to the system is consistent with the earlier results calculated in the framework of extensive statistics \cite{castorina}.  However, by taking higher $q$-values in non-extensive statistics, $c_{s}^{2}$ increases near the limiting temperature as compared to the extensive Boltzmann statistics. 
Also, it is observed that the cut-off effect appears at lower temperature for higher values of ``$q$". This indicates that if there is any temperature fluctuation inside the system then the criticality in the speed of sound shifts towards the lower temperature in the phase diagram. In other words, the phase transition in non-extensive systems is achieved at lesser temperatures as compared to the systems following extensive statistics. Our results leave  open the possibility that there exists a special value of $q$ where a phase transition is no longer present. Taking a mass cut-off of 2.5 GeV in the physical resonance gas, we have studied the $c_{s}^{2}$ as a function of system temperature, for different $q$-values and found that for $q$-values higher than 1.13, all criticality disappears. The present study bears significant implications at the LHC and higher energies, where one expects to study event-by-event temperature fluctuation specifically for the systems which stay away from thermal equilibrium.


\end{document}